\documentclass[12pt]{article}
\usepackage{times}
\usepackage{geometry}
\usepackage[utf8]{inputenc}
\usepackage{enumitem,amssymb}
\usepackage{ragged2e}
\usepackage[round]{natbib} 
\usepackage{indentfirst} 
\usepackage{pifont}
\usepackage{graphicx}
\usepackage{caption}
\usepackage{subcaption}
\newcommand{\cmark}{\ding{51}}%
\newcommand{\done}{\rlap{$\square$}{\raisebox{2pt}{\large\hspace{1pt}\cmark}}%
\hspace{-2.5pt}}

\begin{document}
\raggedright
\huge
Precision measurement of magnetic field from near to far, from fine to large scales in ISM \linebreak
\normalsize

\noindent \textbf{Thematic Areas:} \hspace*{60pt} $\done$ Planetary Systems \hspace*{10pt} $\done$ Star and Planet Formation \hspace*{20pt}\linebreak
\linebreak

\textbf{Principal Author:}

Name: Huirong Yan	
 \linebreak						
Institution:  Deutsches Elektronen-Synchrotron \& University of Potsdam, Germany
 \linebreak
Email: huirong.yan@desy.de
 \linebreak

\textbf{Co-authors:} (names and institutions)

C\'ecile Gry, LAM,  Aix Marseille Univ, CNRS, CNES \linebreak
Francois Boulanger, Laboratoire de Physique de l'ENS, Universit\'e PSL, CNRS \linebreak
Francesco Leone, University of Catania \linebreak

\justify

\textbf{Abstract:}
Magnetic fields have important or dominant effects in many areas of astrophysics, but have been very difficult to quantify. Spectropolarimetry from Ground State Alignment (GSA) has been suggested as a direct tracer of magnetic field in interstellar diffuse medium. Owing to the long life of the atoms on ground states the Larmor
precession in an external magnetic field imprints the direction of the field onto the polarization of absorbing species. This provides a unique tool for studies of sub-gauss magnetic fields using polarimetry of UV, optical and radio
lines. Many spectral lines with strong signals from GSA are in the UV band. By discerning magnetic fields in gas with different dynamical properties, high spectral resolution measurement of spectral polarization will allow the study of 3D magnetic field distribution and interstellar turbulence. GSA provides also a unique chance to map 3D direction of magnetic field on small scales, e.g., disks, where grain alignment is unreliable. The range of objects suitable for studies is extremely wide and includes magnetic fields in the interplanetary medium, in the interstellar medium, and in circumstellar regions as well as diffuse media in extragalactic objects.

\pagebreak

\section{Magnetic field measurement in interstellar medium}

Observational studies of magnetic fields are vital as magnetic fields play a crucial role in various astrophysical processes, including star and planet formation, accretion of matter, transport processes. The existing methods of studying magnetic field have their limitations. Therefore it is important to explore new effects which can bring information about magnetic field. "Ground state alignment" has been identified as an innovative way to determine the magnetic field in diffuse medium. The atoms get aligned in terms of their angular momentum and, as the life-time of the atoms/ions we deal with is long, the alignment
induced by anisotropic radiation is susceptible to very weak magnetic
fields ($1G > B > 10^{-15}G$, \citealt{YL12}), which is exactly the level of the magnetism in diffuse medium, including both ISM and IGM.  

Recent dust polarization measurements by e.g., Planck, have represented a huge step forward in the knowledge of the Galactic magnetic field in terms of sensitivity, sky coverage and statistics. However, they tell us nothing on the distance of the magnetic field and its distribution in the different components or phases. Observing several hundred of hot stars with a S/N of 500 to measure linear polarization from optically thin UV absorption lines, will provide us exclusive information on magnetic field distribution and turbulence properties in different interstellar phases.

Most of the resonance absorption lines are in the {\em UV} domain. A UV band polarimeter with high spectral resolution (R $> 20000$) will thus provide an incomparable opportunity for precision magnetic field measurement, which no other current instruments can offer. Particularly, the high spectral resolution allows simultaneous determination of both velocity and magnetic field, filling the gap of 3D magnetic tomography in ISM, which is so far missing.  The resonance absorption lines appropriate for studying
magnetic fields in diffuse, low column density ($A_V \sim$ few tenths)
neutral clouds in the interstellar medium are NI, OI, SII, MnII, and
FeII. These are all in the ultraviolet.  At higher column densities, the above lines become optically thick,
and lines of lower abundance, as well as excited states of the above
lines become available.

 \begin{figure}[h]
\begin{centering}
 \includegraphics[width=0.45\textwidth]{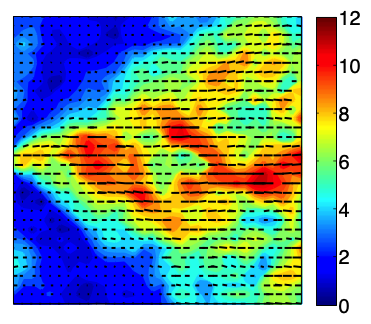}
  \includegraphics[width=0.4\textwidth]{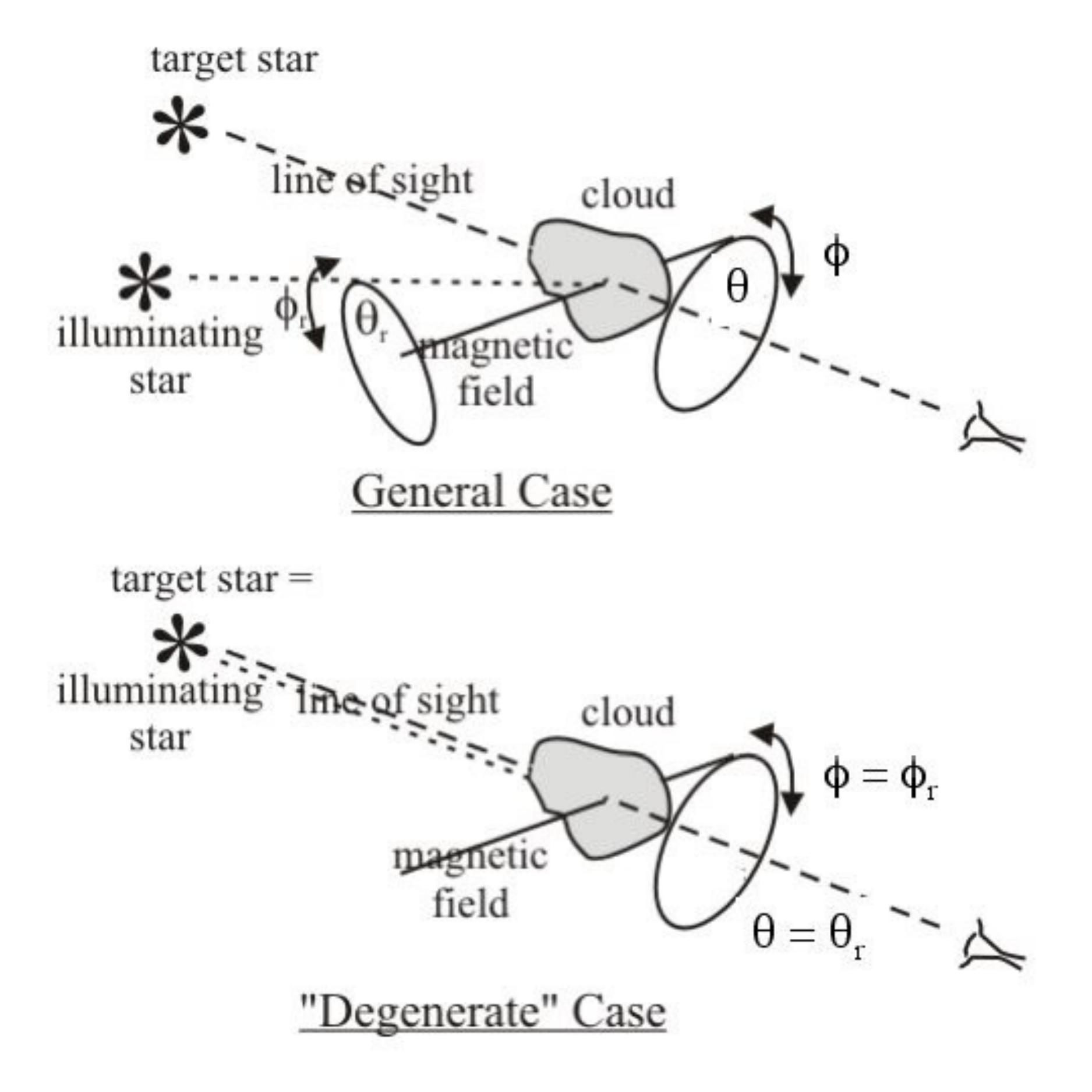}

\caption{\small {\it Left Panel}: Synthetic polarization map of the simulated super-Alfvenic diffuse ISM. The size of the field is 1pc$^2$, with an O/B star located at 0.1pc to the left. (a) The contour color reveals the percentage of polarization induced in the S II 1250A absorption line and the orientation of the bars represents the direction of the polarization. The expected polarization is mostly above 5\%. Even the general interstellar radiation field presents enough anisotropy to align the atoms and create GSA \citep{zhang15}.
{\it Right Panel}: Typical astrophysical environment where 
GSA can happen. A pumping source deposits angular momentum to 
atoms in the direction of radiation and causes differential occupations on 
their ground states. In a magnetized medium where the Larmor precession 
rate $\nu_L$ is larger than the photon arrival rate $R_F$, however, 
atoms are realigned with respect to magnetic field. Observed polarization depends on both $\theta_r$ and 
$\theta$, the angle between the magnetic field and the line of sight. 
In general, there are two situations: ({\it upper part}), the alignment 
is produced by a pumping source while we observe another weak background 
source whose light passes through the aligned medium; ({\it lower part}), 
the background source coincides with the pumping source, in this case, 
$\theta_r=\theta$.
}
  \end{centering}
\label{SIIobs}
\end{figure}


As a first step, with low resolution measurement, 2D magnetic field in the pictorial plane can be easily obtained from the direction of polarization with a 90 degree degeneracy, similar to the case of  grain alignment and Goldreich-Kylafis effect \citep{Goldreich:1981dz}. UV absorption lines are polarized through GSA exclusively. Any polarization, if detected, in absorption lines, would be an exclusive indicator of alignment, and it traces magnetic field 
since no other mechanisms can induce polarization in absorption lines. 
 With high resolution spetropolarimeter, 3D direction of magnetic field can be extrapolated by combining the polarization of two lines or one line polarization with their line intensity ratio, which is influenced by the magnetic field as well \citep[see][]{YLfine, ZYR18}. With the knowledge of degree of polarizations and/or $\theta_r$, the angle between magnetic field and line of sight, the 90 degree degeneracy can be removed.  

\subsection{Large scale magnetic field distribution and turbulence: 3D tomography}

On large scales, mapping of spectropolarimetry will constrain much better the 3D distribution of magnetic field. Interstellar magnetic field is turbulent with velocity and magnetic fluctuations ranging from large injection scales to small dissipation scales.
High resolution spectroscopy and spectropolarimetry combined bring forth a wealth of information on interstellar turbulence. Most magnetic diagnostics render only averaged mean magnetic field on large scales. In this respect GSA fits a unique niche as it reveals small scale structure of magnetic field (see Fig.1{\it left}). Measuing 3D turbulence will shed light on many problems including star formation and cosmic ray, interstellar chemistry, etc.

In highly turbulent environment, we expect magnetic fields to be entangled. Higher resolution UV instrument would be valuable since it naturally reduces line of sight averaging. If the pumping star is along the line of sight, as in the
central star of a reflection nebula, this is the so-called
``degenerate case" (see Fig.1{\it right}), where the
position angle of the polarization gives the 2D magnetic field in the plane of sky, and the degree gives the angle to the
line of sight. In the more general case, an observed cloud might be pumped
from the side, while the positional angle of magnetic field is
available with $90^o$ degeneracy, the derivation of the full magnetic
geometry requires measuring two lines either from the same species or
from two different species.  
  
\subsection{Small scale magnetic field: 3D direction}

On small scales, spectropolarimetry from GSA is an ideal tracer of local magnetic fields. Examples include disks, local bubble and PDR regions. One interesting case is circumstellar disks, for which grain alignment has been found unreliable. In the case of pre-main sequence stars, pumping conditions are similar to those for comets in the Solar System  \citep[see][]{Shangguan2013}: pumping rates on the order of 0.1 - 1 Hz, and realignment for fields greater than 10 -100 mGauss.  Conditions here are apparently conducive to substantial populations in CNO metastable levels above the ground term: \cite{Roberge:2002fk} find strong absorption in the FUV lines (1000 - 1500 Ang)
of OI (1D) and NI and SII (2D), apparently due to dissociation of
common ice molecules in these disks (these are also common in comet
comae).  Since these all have total angular momentum quantum number $>$1, they should be pumped, and
realigned.  This presents the exciting possibility of detecting the
magnetic geometry in circumstellar disks and monitoring them with time. The potential has been clearly revealed by the detection on a binary system where 3D magnetic fields are precisely mapped for the first time via polarization of absorption lines \cite{Zhang_GSA19}. 

\subsection{Polarization of other possible lines}

{\bf Resonance and fluorescence lines} The magnetic realignment diagnostic can also be used in resonant and fluorescent scattering lines.  This is because the
alignment of the ground state is partially transferred to the upper
state in the absorption process \citep{YLhyf}.  In the cases where the direction of optical pumping is known, e.g., in planetary system and circumstellar regions, magnetic realignment exhibit itself in a line polarization whose positional angle is neither perpendicular or parallel to the incident radiation. This deviation depends on the magnetic geometry and the scattering angle.  The degree of polarization also depends on these
two factors. In practice, GSA can be identified by comparing the polarizations from alignable and non-alignable species, which do not trace the magnetic field. There are a number of fluorescent lines in emission nebulae that are potential candidates \citep[see][]{Nordsieck:2008kx}. Reflection nebulae would be an ideal place to test the diagnostic,
since the lack of ionizing flux limits the number of levels being
pumped, and especially since common fluorescent atoms like NI and OI
would not be ionized, eliminating confusing recombination radiation.

{\bf IR/submillimetre transitions within ground state} The alignment on the ground state affects not only the optical (or UV) transitions to the excited state, but also the magnetic dipole transitions within the ground state. The submillimetre lines emitted and absorbed from the aligned medium are polarized in a same fashion as that of the absorption lines, i.e., either parallel or perpendicular to the magnetic field \citep{YLHanle, ZY18}. Current facilities, e.g., SOFIA and ALMA, already have the capability to cover the submillimetre band for the spectral polarimetry observation.

\subsection{Studying magnetic field strength}

GSA is usually by itself not directly sensitive to the magnetic field strength. The exception from this rule is a special
case of pumping photon absorption rate being comparable with the Larmor frequency \citep[see][]{YLHanle,Zhang_GSA19}. However, this should not
preclude the use of GSA for studies of magnetic field. Grain alignment is not sensitive to magnetic field strength either.
This does not prevent polarization arising from aligned grains to be used to study magnetic field strength with the so-called
Chandrasekhar-Fermi technique \citep{Chandrasekhar:1953nx}. In this technique the fluctuations of the magnetic field direction are associated with Alfven perturbations and therefore simultaneously measuring the velocity dispersion using optical/absorption lines arising from the same regions
it is possible to estimate the magnetic field strength. The Chandrasekhar-Fermi technique and its modifications (see also \citealt{Hildebrand:2009zr, Falceta-Goncalves:2008ve,CY16} for
the modifications of the Chandrasekhar-Fermi technique) can be used to find magnetic field strength using GSA. 

The advantage of using spectral lines compared to dust grains is that both polarization and line broadening can be measured from the same lines, making sure that both polarization and line broadening arise from the same volumes. In addition, GSA, unlike grain alignment does not contain ambiguities related to dust
grain shape. Thus, potentially, Chandrasekhar-Fermi technique can be more accurate when used with GSA. 
 
 \section{Synergy of techniques for magnetic field studies}

The GSA and grain alignment complement each other. For instance, measurements of grain alignment in the region where GSA is mapped for a single species
can remove the ambiguities in the magnetic field direction. At the same time, GSA is capable of producing a much more detailed map of magnetic field in the diffuse gas and measure magnetic field direction in the regions where the density
of dust is insufficient to make any reliable measurement of dust polarization. In addition, for interplanetary magnetic field measurements it is important that GSA can measure magnetic fields on time scales much shorter than aligned grains. For the latter, the minimal time scale that they can trace should be larger than their Larmor period.

The synergy exists with other techniques as well. For instance, GSA allows to reveal the 3D direction of magnetic field. This gives the direction of the magnetic field, but not its amplitude. If Zeeman effects allow to get the amplitude of one projected component of magnetic field, this limited input enables GSA to determine the entire 3D vector of magnetic field including its amplitude. The importance of such synergetic measurements is difficult to overestimate.

As astrophysical magnetic fields cover a large range of scales, it is important to have techniques to study magnetic
fields at different scales. For instance, we have discussed the possibility of studying magnetic fields in the interplanetary medium. This can be done without conventional expensive probes by studying polarization of spectral lines. In some cases
spreading of small amounts of sodium or other alignable species can produce detailed magnetic field maps of a particular regions of the interplanetary space, e.g. the Earth magnetosphere.

\section{Summary}

GSA is an important effect the potential of which for magnetic field studies has not been fully
utilized by the astrophysical community. 
The alignment itself is an effect studied well in the laboratory;
the effects arise due to the ability of atoms/ions with fine and hyperfine structure to
get aligned in the ground/metastable states. Due to the long life of the atoms in such states the Larmor
precession in the sub-gauss magnetic field imprints the direction of the field into the polarization of spectral lines ranging from UV, optical to submillimeter and radio lines. UV spectropolarimetry provides an inimitable opportunity since most of the resonant absorption lines are in UV band. The range of objects for studies is extremely wide and includes magnetic fields in the early universe, in the interplanetary medium, in the interstellar medium, in circumstellar regions as well as diffuse media in extragalactic objects. Last but not least, the consequences of the alignment should be taken into account for correct determination of the abundances of alignable species.

\pagebreak



\end{document}